\documentclass[preprint,12pt]{elsarticle}

\usepackage{graphicx}
\usepackage{amssymb}

\usepackage{lineno}
\usepackage{color} 
\usepackage{url} 
\usepackage{array} 
\newcolumntype{C}[1]{>{\centering\arraybackslash}m{#1}}
\usepackage[normalem]{ulem} 
\usepackage{multirow} 
\usepackage{booktabs}

\usepackage{xcolor}

\usepackage{ifthen}
\newboolean{showcomments}
\setboolean{showcomments}{true} 
\ifthenelse{\boolean{showcomments}}
 {
 \newcommand{\claes}[1]{\textcolor{cyan}{{\it [Claes says: #1]}}}
 \newcommand{\austen}[1]{\textcolor{magenta}{{\it [Austen says: #1]}}}
 \newcommand{\othercomment}[1]{\textcolor{blue}{{\it [#1]}}}
 }
 {
 \newcommand{\claes}[1]{}
 \newcommand{\austen}[1]{}
 \newcommand{\othercomment}[1]{}
 }

\journal{Elsevier journal}

\begin{document}


 

\begin{frontmatter}




\title{Recruiting credible participants \\ for field studies in software engineering research}



\author[label1]{Austen Rainer}
\author[label2]{Claes Wohlin}
\address[label1]{Queen's University Belfast, 18 Malone Road, Computer Science Building, BT9 5BN, Belfast, Northern Ireland, UK}
\address[label2]{Blekinge Institute of Technology, SE-371 79 Karlskrona, Sweden}



\begin{abstract}
\textbf{Context}: Software practitioners are a primary provider of information for field studies in software engineering. Research typically recruits practitioners through some kind of sampling. But sampling may not in itself recruit credible participants.\\
\noindent \textbf{Objectives}: To propose and demonstrate a framework for recruiting professional practitioners as credible participants in field studies of software engineering.
\\
\noindent \textbf{Method}: We review existing guidelines, checklists and other advisory sources on recruiting participants for field studies. We develop a framework, partly based on our prior research and on the research of others. We search for and select three exemplar studies (a case study, an interview study and a survey study) and use those to demonstrate the application of the framework.\\
\noindent \textbf{Results}: Whilst existing guidelines etc. recognise the importance of recruiting participants, there is limited guidance on how to recruit the right participants. Our demonstration of the framework with three exemplars shows that at least some members of the research community are aware of the need to carefully recruit participants.\\
\noindent \textbf{Conclusions}: The framework provides a new perspective for thinking about the recruitment of credible practitioners for field studies of software engineering. In particular, the framework identifies a number of characteristics not explicitly addressed by existing guidelines.
\end{abstract}


\end{frontmatter}





\section{Introduction}
\label{section:introduction}

\subsection{Motivation}

Broadly speaking, empirical software engineering research collects data from two sources: artefacts and people. There are many reasons why software engineering (SE) researchers collect data from software practitioners. We consider two. First, many aspects of SE are not observable in, or through, artefacts, e.g., the context in which a software project is progressing. Because the artefacts cannot provide the information needed for the research, the researcher relies on information from people as a kind of substitute or proxy. Second, given human--centric software engineering and behavioural software engineering~\cite{lenberg2015behavioral}, researchers are interested in qualities and experiences of the people themselves, e.g., in their goals, motivation, stress, happiness and values.

There are several research methods for collecting data from people, e.g., interviews, surveys, and focus groups. Fundamental to all of these methods is the need to collect data from software practitioners who are \textit{credible participants for that study}. In a previous paper~\cite{wohlin2021evidence} we defined ``credible evidence''. Drawing on that definition, we define a ``credible participant'' as a person that we trust, believe or rely upon to provide credible evidence. This implies that researchers need to ensure and assure that, for example, a study is using participants who can provide rigorous and valid  information that is relevant to that study.

Guidelines and checklists in SE research tend to discuss participants in terms of populations and samples, e.g., the researcher should define the population of interest and sample from that population, or from a sampling frame. In practice, sampling tends to be based on the practitioner's functional role, perhaps with the additional inclusion of experience or expertise, e.g., experienced software architects, or novice programmers. Also, researchers tend to use a \emph{convenience sample}~\cite{bouraffa2020two}, often with self--selecting participants, e.g., respondents aware of and willing to participate in, for example, a survey.  

In our previous research~\cite{wohlin2021evidence}, we reasoned about \emph{credible evidence}. That research did not address the credibility of information obtained from participants in different empirical studies in SE. Based on our authorship of books on case study research~\cite{runeson2012book} and controlled experiments~\cite{wohlin2012experimentation}, we hypothesised that advice on \emph{recruiting credible participants} for field studies is limited. Our perception was that the level of advice is similar today as it was in 2012, when the latest editions of the books were published. Thus, we decided to investigate if our perception was correct, i.e., that existing guidelines and checklists give insufficient advice on the recruitment of credible participants. For example, sampling based on functional role may not be an effective criterion for selecting credible participants because the functional role may not, of itself, be appropriate for the study. Others, e.g., Lenarduzzi et al.~\cite{lenarduzzi21}, recognise similar challenges with recruiting participants for \textit{experimental} studies in software engineering.

If we are uncertain about the credibility of the participants, on what basis can we have confidence in the reliability, validity and relevance of the information those participants provide? As we discuss in Section~\ref{section:related-work}, adding more participants - increasing the sample size - does not in itself address concerns about reliability, validity and relevance. Such concerns affect all research methods that collect data from participants, and therefore may affect the results of already--completed empirical studies as well as many potential studies in the future

\subsection{Aim of this article}

The aim of this article is to propose and demonstrate a framework for recruiting credible participants in field studies of software engineering. To help us develop and demonstrate the framework, we investigate two research questions:

\begin{description}
    \item[RQ1:] What guidance is currently available on the recruitment of credible participants for field studies of software engineering?
    \item[RQ2:] How might we determine a credible participant?
\end{description}

To answer RQ1, we review published guidelines and checklists in empirical software engineering research, as well as articles providing implicit guidance. We had anticipated conducting a systematic mapping study, or similar, of prior research, however following a review of the guidelines we conclude that a systematic mapping study would not substantively add anything new to our preliminary review.


To answer RQ2, we develop a framework and demonstrate the framework using three exemplar studies.

\subsection{Scope of the article}
\label{subsection:scope-of-article}
To delimit our discussion, we focus this article on the use of participants as providers of information in field studies. Furthermore, we focus on participants as providers of information about software phenomena that are \emph{external} to the participant. The requirement is that the participant can, in principle, perform or observe the phenomena in its real world setting. We include, for example, case studies, interviews and surveys of software development. We exclude empirical studies investigating the internal characteristics of participants, such as their motivation, or (un)happiness~\cite{graziotin2017unhappiness}, or stress. Internal characteristics may, of course, affect the ability of participants to be credible performers or observers of external behaviour. We also exclude empirical studies of controlled situations, the typical example being the experiment. As an example, Lenarduzzi et al.~\cite{lenarduzzi21} are working on a methodology for the selection of participants in software engineering \textit{experiments}.

\subsection{Contribution}

The article makes the following contributions:
\begin{enumerate}
    \item Corroborates our hypothesis, through a review of guidelines and some additional advisory sources, i.e., answering RQ1, that there is limited guidance currently available on the recruitment of credible practitioners. Furthermore, in corroborating this hypothesis, we identify a gap in the empirical software engineering research community’s thinking about recruiting participants for field studies, i.e., researchers tend to think only or primarily in terms of sampling participants in contrast to thinking about recruiting the ``right'' participants.
    \item Proposes a framework for thinking about how to recruit credible participants when collecting data in field studies.
    \item Demonstrates, through application of the framework to three exemplar studies, that existing guidelines, checklists and other advisory sources currently do not address the issue of recruiting the right participants.
    \item Demonstrates, through application of the framework, the value of the framework, i.e., helping researchers to think about recruiting appropriate participants.
\end{enumerate}

\subsection{Structure of the article}

The remainder of this article is organised as follows: Section \ref{section:related-work} reviews relevant prior work; Section~\ref{section:method} explains our research approach for reviewing existing advisory sources, formulating the framework and demonstrating the framework for three exemplar studies; Section~\ref{section:analysis-guidelines} analyses the guidelines and other advisory sources concerning their treatment of participant recruitment for field studies; Section~\ref{section:framework} presents the framework itself; Section~\ref{section:reference-examples} discusses the three exemplar studies (i.e., a case study, an interview study and a survey study) used for comparison with the framework; and, finally, Section~\ref{section:conclusion} concludes. 

\section{Related work}
\label{section:related-work}

For our review of related work, we begin with a discussion of the flow of information through a SE research study. This discussion provides a context for thinking about the credibility of participants as providers of information into that flow. We then consider the problem of participant accuracy, referring to two articles~\cite{bernard1984problem,kronenfeld1972toward} published in anthropology. We consider participant accuracy because it can be used as an indicator of participant credibility and because the two articles demonstrate significant challenges with participant accuracy. Next, we discuss sampling in SE research. We distinguish between the item of interest in a sample and the participant as a research instrument to study that item of interest. We use a published field study~\cite{curtis1988field} as an example to illustrate the difference between items of interest and participants. Because participants have a different status, as research instruments rather than subjects in a sample, we then consider participants as key informants. Finally, we connect our preceding discussions to the R\textsuperscript{3} model~\cite{falessi2018empirical} of participant experience, and finally summarise the review of related work.

\subsection{The flow of information in SE research}
\label{subsection:information-flow}

\begin{figure}
    \centering
    \includegraphics[scale=0.4]{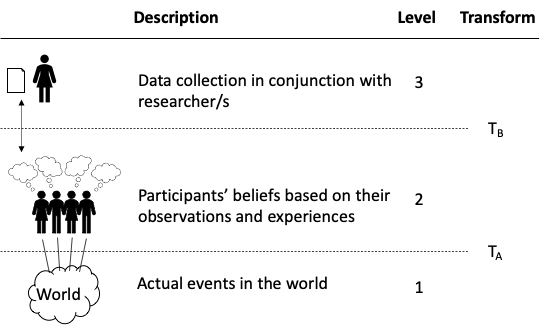}
    \caption{The source and flow of information in research (derived from Karlstr{\"o}m and Runeson \cite{karlstrom2006integrating})}
    \label{figure:generalised-karlstrom-and-runeson-figure}
\end{figure}

In their study of two large, independent software projects, Karlstr{\"o}m and Runeson \cite{karlstrom2006integrating} present a model of the flow of information in the research process. We present a version of Karlstr{\"o}m and Runeson's \cite{karlstrom2006integrating} model of the flow of information in Figure \ref{figure:generalised-karlstrom-and-runeson-figure}. We have simplified the model to make it more relevant to our article. Information about the world is based on the participants' perceptions of the actual world or, as Bernard et al.~\cite{bernard1984problem} recognises, based on cultural norms about the world. The information is transformed (cf. T\textsubscript{A} and T\textsubscript{B}, in Figure \ref{figure:generalised-karlstrom-and-runeson-figure}) as it flows through the research process. This transformation may affect, amongst other qualities, the validity, reliability and relevance of the information.

Because Karlstr{\"o}m and Runeson \cite{karlstrom2006integrating} conduct a case study (of two cases), the participants are recruited in relation to specific, identifiable situations in the world. In Karlstr{\"o}m and Runeson's \cite{karlstrom2006integrating} study, the \textit{relationship} of the participants to the set of events in the world is therefore relatively well known. A consequence is that the researchers can have more confidence in the information provided by the participants, e.g., because the researchers can assess the relationship. Furthermore, Karlstr{\"o}m and Runeson~\cite{karlstrom2006integrating} also interview the participants, providing the opportunity (again at least in principle) to clarify, challenge, etc. the information provided by those participants. In other words, the researchers know, or can know, something about the source of the information and about the nature of the transformations between levels of information (at T\textsubscript{A} and T\textsubscript{B}, in Figure \ref{figure:generalised-karlstrom-and-runeson-figure}), such as the information--selection decisions being made at levels 2 and 3.

Circumstances can be very different for interview studies and survey studies. The researcher may have less influence or control on the participants recruited, e.g. using a convenience sample. Because of the nature of the interview study, the researcher may know something about the nature of transformations, may be able to influence those transformations (e.g., to gather information based on actual experience rather than cultural norms), and may be able to know something about the information-selection occurring at levels 2 and 3. But because of the nature of the survey study, the researcher has limited, if any, influence or control on the situations in the world to which the participants refer. The researcher has limited opportunity to influence the information provided by the participants, but nevertheless still has some opportunity. The researcher knows little about the nature of transformations at T\textsubscript{A} and T\textsubscript{B}, or about the information selected at level 2 and the information shared from level 2 to level 3.

The quality of information flowing through the research is therefore fundamentally dependent on the quality of information provided at the source of the process, i.e., the information provided by the participant.

\subsection{The problem of participant accuracy}
\label{section:accuracy}

Bernard et al.~\cite{bernard1984problem} consider the problem of participant accuracy and the validity of retrospective data. They observe that anthropology researchers often ask participants to provide data on, as examples, their (i.e., the participant's) behaviour, on the behaviour of others, on sequences of events, and on economic and environmental conditions. Bernard et al.~\cite{bernard1984problem} identify three areas in which, at the time of their review, information accuracy had been moderately well studied - i.e., recall of childcare behaviour, recall of health seeking behaviour, and recall of communication and social interaction - together with a fourth area that attempted to deal constructively with the problem of participant accuracy.

Given their review of prior work, Bernard et al.~\cite{bernard1984problem} present many examples, arguments and conclusions. For conciseness, we present one example here and use that example as the basis of our discussion of their work. The example is drawn from an experiment conducted by Kronenfield et al.~\cite{kronenfeld1972toward} in which informants leaving a restaurant were asked to report on what the waiters and waitresses were wearing, as well as the music being played. Kronenfield et al.~\cite{kronenfeld1972toward} found much higher agreement about what waiters were wearing than what waitresses were wearing. This was despite the fact that none of the restaurants in question had waiters. Similarly, they found that informants provided greater detail about the kind of music that was playing in restaurants that were, in fact, not playing music. This example, together with Kronenfield et al.'s~\cite{kronenfeld1972toward} interpretation of the results, and other work that Bernard et al. reviewed, leads Bernard et al.~\cite{bernard1984problem} to suggest the following:

\begin{enumerate}
    \item Participants who have \emph{actually} observed an event or circumstance are able to report \emph{parts} of the actual event or circumstance. They can only report parts because, for example, they can only recall part of the actual experience.
    \item By contrast, those who have \emph{not} observed an event or circumstance start from cultural norms. They are able to provide ``\dots rich descriptions, unencumbered by partial memories and working from complex normative wholes, based on many experiences over a lifetime.'' (\cite{bernard1984problem}, p. 510). In other words, participants who have not actually experienced the event, infer (not necessarily consciously) apparently more complete information about the event.
 
    \item Interviewing many inaccurate participants will not solve the accuracy/ validity problem, and will, as a consequence, also not produce relevant findings.
\end{enumerate}

Bernard et al.'s~\cite{bernard1984problem} observation about cultural norms is demonstrably present in software engineering. For example, Rainer et al.~\cite{rainer2003persuading} found that practitioners prefer local opinion over other sources of knowledge. And Devanbu et al.~\cite{devanbu2016belief} conducted a large survey of Microsoft employees ($n=564$) finding that developers' beliefs are based primarily on their personal experience and then, second, on their peers' opinions. By contrast, research papers were ranked fifth out of six. Thus, when practitioners cannot rely on their own experience they appear to first turn to others, i.e., to the source of cultural norms. This raises a serious implication for SE research, i.e., consistency of responses across a sample of apparently independent participants may be explained by cultural norms rather than by a consistent behaviour in the phenomenon of interest. 

\subsection{Sampling}
\label{subsection:sampling}

In their review of prior work, Bernard et al.~\cite{bernard1984problem} write, ``It is time we determined whether key informants [participants] are really better than a \emph{representative survey}\dots'' (\cite{bernard1984problem}, p. 513; emphasis added here).

In a recent article, Baltes and Ralph~\cite{baltes2020sampling} provide a primer on sampling in SE research. They define sampling as the process of selecting a smaller group of items to study, the \emph{sample}, from a larger group of items of interest, the \emph{population}. A \emph{sampling frame} is the available population list from which a sample can be actually drawn.

Baltes and Ralph~\cite{baltes2020sampling} identify two problems with sampling frames: first, for many software engineering phenomena there is no suitable sampling frame from which to draw a sample; second, some software engineering studies adopt poorly understood sampling strategies such as random sampling from a non--representative surrogate population. Baltes and Ralph~\cite{baltes2020sampling} also write, ``For our purposes, \emph{representativeness} is the degree to which a sample's properties (of interest) resemble those of a target population.'' (\cite{baltes2020sampling}; emphasis in original).

We can \emph{model} a participant as a kind of research instrument - a lens - with which, or through which, we can study software practice. For example, in their field study of software design for large systems, Curtis et al.~\cite{curtis1988field} interviewed 97 participants across 17 projects in 9 companies. In our terminology, Curtis et al.~\cite{curtis1988field} used 97 research instruments to observe the behaviour of software development at five levels of behaviour: the individual, the team, the project, the company and the business milieu. To clarify, the individual level did not refer to the 97 participants each introspectively studying themselves, but rather the 97 participants provided information on the behaviour at the individual level in software development.

When modelling the participant as a research instrument, the participant is not the item of interest, but is instead a means to study the item of interest. Taking this perspective, the item of interest in Curtis et al.'s~\cite{curtis1988field} study is the software project designing a large system. The 17 projects in 9 companies are therefore the sample of items of interest, and the 97 participants are instruments to study that sample. Furthermore, at least in principle, these 97 participants are selected as the more credible participants for providing information on the 17 items of interest.

\subsection{Key informants}
\label{section:key-informants}

As noted in the preceding subsection, the 97 participants in Curtis et al.'s~\cite{curtis1988field} field study were, at least in principle, the more credible participants for providing information. Marshall~\cite{marshall1996key} defines a key informant as an expert source of information. The principal of a key informant is that the informant can provide more reliable, valid and relevant information than a sample of participants. In other words, a key informant can be a more credible participant than a sample. This relates back to Bernard et al.'s~\cite{bernard1984problem} review, e.g., that more participants will not in itself solve the accuracy problem, that researchers seek participants who can provide information on the basis of actual experience, and we seek to avoid information based on cultural norms.

Marshall identifies five characteristics of the ideal key informant, summarised here in Table~\ref{table:key-informants}. All five characteristics contribute to participant credibility.

\begin{table}[!ht]
    \small
    \caption{Characteristics of ideal key informant (from \cite{marshall1996key})}
    \label{table:key-informants}
    \centering
    \begin{tabular}{| l | p{12cm} | }
        \toprule
        \textbf{\#} & \textbf{Description}\\
        \midrule
        C1 & \textbf{Role in community}. Their professional role in their peer community should expose them to the kind of information being sought by the researcher.\\
        C2 & \textbf{Knowledge}. In addition to having access to the information desired, the informant should have absorbed the information meaningfully.\\
        C3 & \textbf{Willingness}. The informant should be willing to communicate their knowledge to the interviewer and to cooperate as fully as possible.\\
        C4 & \textbf{Communicability}. They should be able to communicate their knowledge in a manner that is intelligible to the interviewer.\\
        C5 & \textbf{Impartiality}. The key informant should be objective and unbiased. Any relevant biases should be known by the interviewer, e.g., the key informant declares a bias or the interviewer can determine this from other sources.\\
        \bottomrule
    \end{tabular}
\end{table}

\subsection{The R\textsuperscript{3} model}
\label{subsection:Rmodel}

Falessi et al.~\cite{falessi2018empirical} propose the R\textsuperscript{3} model comprising three elements of a participant's experience: Real, Relevant and Recent. The R\textsuperscript{3} model was formulated for experiments however it has relevance to our discussion, i.e., a participant with Real, Relevant and Recent experience is more likely to satisfy the requirements of a credible participant.
We summarise the R\textsuperscript{3} model here and then reformulate it in Section \ref{section:framework} to align with our proposed framework.

The three elements of the R\textsuperscript{3} model are: 
\begin{itemize}
    \item That the subject has \emph{real experience} of software engineering situations. It is not possible to provide a formal definition for ``real experience''. Broadly speaking, ``real experience'' refers to experience of situations of real--world software practice that are, in general, of interest to the research.
    \item That the subject has \emph{relevant experience}. Relevance here refers to the fit between the situation and the research objective. The characteristic of relevance becomes more significant for a type of informant we discuss later in this article, i.e., the \emph{advisor}.
    \item That the subject has \emph{recent experience}, or more precisely \emph{timely experience}, e.g., typically that the situation has been experienced recently by the participant, relative to the focus of the research.
\end{itemize}

All three of the above elements need to be tailored, by the researcher, to the specific needs of the respective research. The researcher will also need to assess each participant against each of the (possibly tailored) elements.

\subsection{Summary}

SE research needs to recruit more participants, and more credible participants, for our field studies of software practice. These participants need to be systematically and carefully described to enable other researchers to judge the credibility of both the participants and the information they provide. Such information is the foundation of field studies that rely on practitioners, e.g., case study, interview, survey and focus group. Information from practitioners that is inaccurate, or information for which its accuracy is highly uncertain, fundamentally undermines the credibility of our subsequent research findings, regardless of the reliability of the research process itself. Rephrasing Bernard et al.~\cite{bernard1984problem}, many studies all reporting inaccurate findings will not solve the credibility problem. 

Treating the recruitment of participants as a matter of better sampling, so as to produce a more representative sample, is potentially mis--framing the problem. This is because participants themselves may not be the item of \emph{research} interest in a field study. In our SE field studies, we are often interested in what participants can tell us \emph{about} the item of research interest, e.g., what interviewees can say about the layers of behaviour of software projects designing large software systems~\cite{curtis1988field}. The distinction between the participant and the item of interest is also important because of the effects of cultural norms, e.g., a participant may provide a culturally--normative response because the participant has not reliably observed or performed the item of interest and cannot provide valid information on that item of interest. Again, many studies all reporting culturally-normative findings will not help us build a body of knowledge of software practice.

As a step towards addressing these concerns, we first review existing guidelines, then propose a framework, and subsequently demonstrate the framework with three exemplars.

\section{Research approach}
\label{section:method}

As noted in Section \ref{section:introduction}, we previously considered \emph{credible evidence}~\cite{wohlin2021evidence} however we did not, in that publication, address the credibility of participants as a source of information for field studies in SE. In SE, field studies often involve some kind of collaboration between industry and academia. Also, based on our authorship of prior work on research methodology~\cite{runeson2012book, wohlin2012experimentation}, we perceived the need for more support to researchers on how to identify and select more credible participants as a source of information for field studies. To ensure that our perception was correct, we review (in Section~\ref{section:analysis-guidelines}) guidelines for different types of field study. Based on our review of existing guidelines, we then propose a framework for researchers (in Section~\ref{section:framework}) and then demonstrate that framework (in Section~\ref{section:reference-examples}). In the current section, we explain our approaches to the review  of guidelines, the development of the framework and the demonstration of the framework.

\subsection{Reviewing guidelines and other advisory sources}
\label{section:reviewingguidelines}

To investigate RQ1, i.e., ``What guidance is currently available on the recruitment of credible participants for field studies of software engineering?'', we reviewed published guidelines and recommendations. In doing so, we also assessed the correctness of our hypothesis that published guidelines do not cover the aspects we perceive are missing. The lack of attention in the guidelines etc. further motivates the need for a framework to determine credible participants in different forms of field study.

To identify appropriate guidelines and other sources of advice, the two authors independently searched for appropriate articles, primarily using Google Scholar. We use Google Scholar because we seek a sufficient coverage of guidelines and not an exhaustive coverage; we are not, for example, attempting a systematic review. We also, of course, had prior experience of some guidelines, e.g., the case study guidelines by Runeson and H{\"o}st~\cite{runeson2009guidelines}. We shared the suggested guidelines, and discussed them in online meetings, arriving at consensus on the advisory sources to consider.

We prioritised guidelines and recommendations that focused on field studies, or empirical studies in general, and that were published \emph{after} 2009, to be consistent with when the case study guidelines were published. Thus we excluded guidelines, such as Lenarduzzi et al.'s~\cite{lenarduzzi21}, that focused on experimental studies. We had difficulties finding guidelines on interviews, discussed below.

In total we identified six initial sources. These are listed in the upper part of Table~\ref{table:review-advisory-sources} and briefly summarised as follows. 

\begin{description}
    \item{\textbf{Case study:}} Research guidelines for conducting case study research within software engineering were first published as a checklist, by Höst and Runeson~\cite{host2007checklists}, at a conference in 2007. Runeson and Höst then published a more extensive set of guidelines as a journal article in 2009~\cite{runeson2009guidelines} and then as a book~\cite{runeson2012book} in 2012. Verner et al.~\cite{verner2009guidelines} published guidelines for industrial case studies in software engineering at a conference in 2009.

    \item{\textbf{Interview study:}} We did not find guidelines for interview studies in software engineering that are comparable to the case study guidelines or to the survey study guidelines (discussed next). We therefore had to relax our requirement for post--2009 publication for these interview ``guidelines''. We also chose two articles. The first article, by Strandberg~\cite{Strandberg2019}, provides advice concerning ethics in interview studies. The second article, by Hove and Anda~\cite{hove2005experiences}, shares their experiences of conducting interview studies in software engineering.

    \item{\textbf{Survey:}} Empirically evaluated survey guidelines are published by Molléri et al.~\cite{molleri2020empirically} in 2020. Their article synthesises a range of previous guidelines on survey studies, e.g.,~\cite{molleri2019cerse,molleri2016survey}.

    \item{\textbf{Participant recruitment:}} Salleh et al.'s~\cite{salleh2018recruitment} article does not present \emph{guidelines} but rather, in a way that is similar to Hove and Anda~\cite{hove2005experiences}, Salleh et al.~\cite{salleh2018recruitment} share their experiences of conducting research in industrial contexts.
\end{description}

Furthermore, we identified four additional sources for reviewing the selection of participants in different types of field study. The four additional sources are listed in the lower part of Table~\ref{table:review-advisory-sources}. Again, we needed to relax our requirement for post--2009 publication. The four additional advisory sources are:

\begin{description}
    \item{\textbf{Focus groups:}} Kontio et al.'s~\cite{kontio2008focus} book chapter, published in 2008, provides guidelines on the use of focus groups in SE.
    \item{\textbf{Preliminary guidelines:}} Kitchenham et al.~\cite{kitchenham02guidelines} published probably the first set of guidelines on empirical studies in SE, in 2002.
    \item{\textbf{Ethnography:}} Zhang et al.~\cite{zhang2019ethnographic} published guidelines for ethnographic studies in 2019.
    \item{\textbf{Empirical standards:}} Ralph, in conjunction with ACM SIGSOFT, are developing standards for empirical studies in SE~\cite{ACM20} . These standards were first published in 2021. 
\end{description}

For all ten articles listed in Table~\ref{table:review-advisory-sources}, we downloaded PDF copies of each article and searched each PDF for explicit guidance on participant selection. The objective of the searches was to identify formulations in the guidelines and advisory sources in relation to advice concerning recruitment of participants for field studies. We were particularly looking for concrete advice, i.e., beyond general statements concerning the importance of recruiting representative participants, e.g., through sampling. To do this, we searched for nine stemmed words related to the persons in a study and to the activity of recruiting such persons. We searched for the following stemmed words:
\begin{itemize}
    \item[] For person: \texttt{subject*}, \texttt{partici*}, \texttt{respond*} and \texttt{contribu*}
    \item[] For activity: \texttt{select*}, \texttt{identif*}, \texttt{sampl*}, \texttt{find*} and \texttt{recruit*}
\end{itemize}

We discuss our analysis of the ten articles in Section~\ref{section:analysis-guidelines}.

\begin{table}[!ht]
    \small
    \caption{Publications selected for reviewing advice on participant recruitment.}
    \label{table:review-advisory-sources}
    \centering
    \begin{tabular}{| l | l | l |}
        \toprule
        \textbf{Ref} & \textbf{Year} & \textbf{Description}\\
        \midrule
        \midrule
        \multicolumn{3}{| p{7cm} |}{Review existing advisory sources}\\
        \cite{runeson2009guidelines} & 2009 & Case study guidelines by Runeson \& H{\"o}st \\
        \cite{verner2009guidelines} & 2009 & Case study guidelines by Verner et al. \\
        \cite{Strandberg2019} & 2019 & Ethical interviews\\
        \cite{hove2005experiences} & 2005 & Semi-structured interviews\\
        \cite{molleri2020empirically} & 2020 & Survey guidelines\\
        \cite{salleh2018recruitment} & 2018 & Recruiting participants\\
        \midrule
        \multicolumn{3}{| p{9cm} |}{Further review of advisory sources}\\
        \cite{kontio2008focus} & 2008 &  Focus group guidelines\\
        \cite{kitchenham02guidelines} & 2002 &  Empirical studies guidelines\\
        \cite{zhang2019ethnographic} & 2019 & Ethnographic guidelines\\
        \cite{ACM20} & 2020 & Empirical standards\\
        \bottomrule
    \end{tabular}
\end{table}

\subsection{Development of the framework}
\label{subsection:framework-development}

We developed a framework to address the second research question, RQ2, i.e., ``How might we determine a credible participant?'' We used a dialectic process to develop the framework, i.e., the first author devised, and subsequently revised, the framework, and the second author independently reviewed the latest version, providing feedback which lead to subsequent revisions. The dialect process was informed by our respective prior research experience, particularly the books~\cite{runeson2012book, wohlin2012experimentation}, as well as the elements identified in Section~\ref{section:related-work}, e.g., information flow, participant accuracy, the characteristics of key informants, and the R\textsuperscript{3} model~\cite{falessi2018empirical}. Being highly--cited, a number of these sources -- specifically, Bernard et al.,~\cite{bernard1984problem}, Marshall~\cite{marshall1996key}, and our two co-authored books~\cite{runeson2012book, wohlin2012experimentation} -- are established in their respective fields of research. The nature of the dialectic process means that the framework is not \textit{deduced from} these sources and elements but rather \textit{created with} them. The review of guidelines, in Section~\ref{section:analysis-guidelines}, and the demonstration with three exemplars, in Section~\ref{section:reference-examples}, both act as a kind of validation of the framework.

Furthermore, the framework progressed through five revisions before we moved to our demonstration of the framework. After completing our demonstrations, we returned to refine the \emph{presentation} of the framework, i.e., we did not change the content of the framework but simplified the way in which it is presented and described.

\subsection{Demonstrating the framework with exemplar papers}

Having developed the framework, we wanted to confirm whether the elements of the framework can be found in at least some published field studies. Our objective here is not to assess the prevalence of the framework's elements in prior research; rather, we want to simply demonstrate that these elements are considered relevant in at least some of the articles that have good descriptions of participants and their recruitment.

To select primary studies for demonstration, both authors independently searched for candidate articles to consider. We used the following search heuristics:
\begin{itemize}
    \item We used Google Scholar for the searches. We do not need to use multiple search engines because we are not attempting an exhaustive search. We simply need to search a large enough space of academic publications.
    \item We prioritised the more highly cited articles, for two reasons: first, we assumed that the more highly cited articles were more likely to have valuable information in them; second, that the research community would have a greater awareness of these articles and therefore our demonstration would have more obvious relevance to the community.
    \item We prioritised articles having better descriptions of the participants in the studies. Better descriptions would help us more easily find information about the framework's elements. By contrast, in an article that reported less information about participants, the study itself may have considered the elements, but simply not reported them.
\end{itemize}

We then independently read the list of candidate articles and discussed them in online meetings, with the final selection made on the basis of unanimous agreement. During our final selection we considered the following criteria:
\begin{itemize}
    \item Articles published before 2009, since 2009 is the year that both Runeson and H{\"o}st~\cite{runeson2009guidelines} and Verner et al.~\cite{verner2009guidelines} published their guidelines on case study research in SE.
    \item Articles reporting primary studies.
    \item Articles that, taken together, would provide coverage of case study, interview study and survey study.
    \item Articles that, taken together, would provide coverage across journals and conferences.
\end{itemize}

Table~\ref{table:assessment-of-framework} lists the three articles we selected. We discuss these articles in Section~\ref{section:reference-examples}. Table \ref{table:assessment-of-framework} reports only the articles finally selected for the comparison. Many other articles were considered.

\begin{table}[!ht]
    \small
    \caption{Publications selected for demonstrating the framework.}
    \label{table:assessment-of-framework}
    \centering
    \begin{tabular}{| l | l | l |}
        \toprule
        \textbf{Ref} & \textbf{Year} & \textbf{Description}\\
        \midrule
        \multicolumn{3}{| p{7cm} |}{Assess framework against primary studies}\\
       \cite{freimut05} & 2005 & Case study paper\\
        \cite{singer1998practices} & 1998 & Interview paper\\
        \cite{Vredenburg02} & 2002 & Survey paper\\
        \bottomrule
    \end{tabular}
\end{table}

\section{Analysis of published guidelines and other advisory sources}
\label{section:analysis-guidelines}

As discussed in Section~\ref{section:reviewingguidelines}, we searched the PDF files of the ten guidelines and advisory sources. The outcome is summarised in Table~\ref{table:search-counts-guidelines}. The table includes the number of occurrences of the stemmed words listed in Section~\ref{section:reviewingguidelines}, together with the number of relevant quotes found in the respective article, and the frequency of those relevant quotes in relation to the total number of occurrences of the stemmed words.

\begin{table*}[htp]
    \small
    \caption{Summary counts of searches of stemmed words.}
    \label{table:search-counts-guidelines}
    \centering
    \begin{tabular}{| l | l | c | c | c |}
        \toprule
        & \textbf{Research} & \textbf{Occurrence} & \multicolumn{2}{| c |}{\textbf{Quote}}\\
        \textbf{Article} & \textbf{method} & \textbf{\textit{f}} & \textbf{\textit{f}} & \textbf{\%}\\
        \midrule
        \multicolumn{5}{| l |}{\textit{Initial selection of six advisory sources}}\\
        Runeson~\cite{runeson2009guidelines} & Case study & 125 & 8 & 7\\
        Verner~\cite{verner2009guidelines} & Case study & 87 & 8 & 9\\
        Strandberg~\cite{Strandberg2019} & Interviews & 101 & 3 & 3\\
        Hove~\cite{hove2005experiences} & Interviews & 78 & 4 & 5\\
        Molléri \cite{molleri2020empirically} & Survey & 312 & 6 & 2\\
        Salleh~\cite{salleh2018recruitment} & General & 293 & 8 & 3\\
        \midrule
        \multicolumn{5}{| l |}{\textit{Further selection of four advisory sources}}\\
        Kontio~\cite{kontio2008focus} & Focus group & 138 & 8 & 6\\
        Kitchenham \cite{kitchenham02guidelines} & General & 114 & 4 & 4\\
        Zhang~\cite{zhang2019ethnographic} & Ethnography & 116 & 1 & 1\\
        Ralph~\cite{ACM20} & Standards & 45 & 1 & 2\\
        \bottomrule
    \end{tabular}
\end{table*}

Furthermore, we summarise the relevant quotes that were found in the ten articles, in Table~\ref{table:review-of-guidelines-1} and Table~\ref{table:review-of-guidelines-2}. These summaries provide a sense of the focus and coverage of the guidelines, i.e., that existing guidelines are quite ``light'' in their coverage of participant recruitment.

\begin{table*}[htp]
    \small
    \caption{Summary of the analysis of the initial six advisory sources.}
    \label{table:review-of-guidelines-1}
    \centering
    \begin{tabular}{| l | p{11.5cm} |}
        \toprule
        \textbf{Ref.} & \textbf{Summary}\\
        \midrule
        \cite{runeson2009guidelines} & There should be a rationale for the selection of participants. Participants should give informed consent. Participants should be selected for diversity rather than similarity. Participants are not selected for statistical representation. With small samples, participants may be identifiable.\\ 
        
        \cite{verner2009guidelines} & Additional participants may be selected through recommendations during interviews. Citing~\cite{stake1995art}, they selected participants with first-hand experience. Overall, the guidelines focus on selecting sites rather than individuals.\\
        
        \cite{Strandberg2019} & Paper focuses on ethical concerns relating to participants, e.g., identifying places and settings make participants more easily identifiable. There is no guidance on recruiting participants.\\
        
        \cite{hove2005experiences} & It is necessary, and also probably requires a lot of effort, to select participants carefully. Participants should have free choice to participate, e.g., not influenced by their managers. Participants may drop-out, impacting the study. Recruitment of participants should be reported.\\

        \cite{molleri2020empirically} & The authors highlight the importance of a sampling plan, including types of sampling. Furthermore, the guidelines describe the need for anonymity and confidentiality, as well as usability and willingness to participate.\\

        \cite{salleh2018recruitment} & The authors discuss how to recruit industry participants in general, e.g., carefully crafting a call for participation so as to avoid a ``spam effect'', and snowballing through word-of-mouth approaches, such as asking managers. They recognise the need to make specific participants requirements clear in the recruitment process, to ensure there is some benefit to the participants in doing the survey, and to collect data from participants to ensure appropriate sampling.\\
        \bottomrule
    \end{tabular}
\end{table*}

For the different guidelines and advisory sources, they may be summarised as follows.

The guidelines by Runeson and Höst~\cite{runeson2009guidelines} highlight the importance of recruiting suitable participants in relation to the objective of the case study, however the guidelines do not provide guidance on how to make an informed decision concerning participant recruitment. 
Verner et al.~\cite{verner2009guidelines} provide case study guidelines for, as they call it, industry-based studies in software engineering. When it comes to selecting participants, they touch on the subject when providing an example concerning the scope of the case study. However, the guidelines by Verner et al. do not further address selecting credible participants except for mentioning the importance of determining the sampling strategy.

Strandberg~\cite{Strandberg2019} highlights the need to know who the stakeholders are, and to be able to consider the potential benefit and harm that may arise from the research.  Strandberg does not, however, discuss how to identify appropriate participants as interviewees. 

Hove and Anda~\cite{hove2005experiences} highlight the need to select or recruit participants carefully. However, Hove and Anda do not provide experiences concerning the challenges of selecting suitable participants in an interview study.

In the survey guidelines, Molléri et al.~\cite{molleri2020empirically} stress that we should identify and select participants based on characteristics, however their article does not provide support concerning what constitutes essential characteristics when recruiting participants for a survey.


Salleh et al.~\cite{salleh2018recruitment} highlight that it is essential that specific requirements on the participants need to be conveyed to the industrial collaborator. However, Salleh et al.~\cite{salleh2018recruitment} do not provide further details concerning what may make a participant suitable for participation in the research.

\begin{table*}[htp]
    \small
    \caption{Summary of analysis of the additional four advisory sources.}
    \label{table:review-of-guidelines-2}
    \centering
    \begin{tabular}{| l | p{11.5cm} |}
        \toprule
        \textbf{Ref.} & \textbf{Summary}\\
        \midrule
        \cite{kontio2008focus} & The authors highlight the recruitment of representative, insightful and motivated participants. The interactive nature of focus groups can enrich the information collected. Participants for focus groups should be carefully selected to mitigate threats, e.g., recruit participants of equal expertise.\\

        \cite{kitchenham02guidelines} & The authors stress the importance of sampling from a \textit{defined} population to be able to draw conclusions from the study. They discuss dropouts. They highlight the necessity of tracking the characteristics of the participants to be able to determine the effects of dropouts.\\

        \cite{zhang2019ethnographic} & The authors highlight the need to make an informed decision of whom to include in the study.\\

        \cite{ACM20} & The authors highlight the importance of ensuring that the sample is representative of the intended population.\\

        \bottomrule
    \end{tabular}
\end{table*}


Kontio et al.~\cite{kontio2008focus} suggest using purposive sampling, i.e., participants are selected based on their characteristics in relation to the topic of the focus group session.

Kitchenham et al.~\cite{kitchenham02guidelines} provide more generic guidance on empirical studies, though their advice is more focused on controlled experiments and statistical analysis. They argue that subjects should be representative of the population. Their preliminary guidelines do not, however, discuss any specific desirable characteristics concerning the participants.


Turning to more recent guidelines, Zhang et al.~\cite{zhang2019ethnographic} present a critical review and checklist for conducting ethnographic studies in software engineering. The guidelines do not provide guidance on how to assess the credibility of the practitioners being studied.




Finally, the empirical standards~\cite{ACM20} only mention participants in relation to quantitative studies, and do not discuss participants in more qualitative studies such as addressed here.

Overall, none of the ten guidelines and other advisory sources summarised in Table~\ref{table:review-of-guidelines-1} and Table~\ref{table:review-of-guidelines-2} provide actionable advice on \emph{how} to determine the credibility of prospective participants. Thus, the proposed framework, presented in Section~\ref{section:framework} and demonstrated in Section~\ref{section:reference-examples}, complements existing guidelines on conducting field studies, particularly in relation to recruiting participants.

\section{Formulating a framework about credible practitioners}
\label{section:framework}

In this section, we present and discuss our framework for thinking about credible participants and the quality of information they can provide to a field study. We first introduce and discuss several components of the framework, and then concisely present the framework in Section~\ref{section:framework-summary} and Table~\ref{table:summary-of-framework}. We briefly described, in Section~\ref{subsection:framework-development}, how the framework was developed.

For the components, we begin with a simple model of the research process as a reference. We then re-consider the sampling of participants for empirical studies, re-framing this as a problem of recruiting credible participants. Then we introduce elements of the framework -- i.e., the three participant roles, characteristics affecting the quality of information, and demographics -- before concisely presenting the framework in Section~\ref{section:framework-summary}.

Following the presentation of the framework, we describe additional considerations that are beyond the scope of the current paper, present a simple example of the application of the framework, and summarise the contribution of the framework. In Section~\ref{section:reference-examples} we demonstrate the framework.

\subsection{A model of the research process}

\begin{figure}[ht]
    \centering
    \includegraphics[scale=0.6]{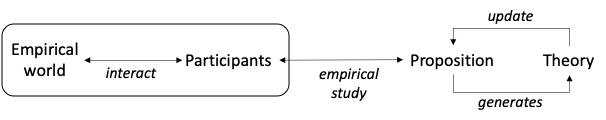}
    \caption{A simple reference model of the research process for field studies in SE}
    \label{figure:research-model}
\end{figure}

Figure \ref{figure:research-model} presents a simple model of the research process for field studies in SE. The model is intended to be used as a reference for the subsequent discussion. In the model, a theory of some kind provides the grounds for a proposition. The proposition is studied empirically.

As already noted, the reference model is a simplification. For example, grounded theories are generated bottom--up from the empirical world. As another example, the propositions of the model may be hypotheses, research questions, or other kinds of testable or empirically investigatable statements.

For empirical software engineering research, many aspects of software engineering practice can be studied directly, e.g., source code, however many other aspects of software engineering practice can only be studied indirectly, e.g., through engaging with software practitioners who themselves interact with the empirical world of software engineering. In the model, practitioners provide information about the empirical world to the researchers as part of an empirical study. As discussed in Section~\ref{section:related-work}, these practitioners may therefore be understood as research instruments. The information that practitioners provide to researchers is broadly of two types: facts that \emph{describe} some aspect of a \emph{specific} software engineering situation, and beliefs about practice that may be specific to a situation or be generalised to more than one situation. Participant demographics may be understood as factual information, however our focus here is on both the factual information and the beliefs that participants provide about the phenomenon of interest.

Again for simplicity, we assume that theories are constructed from four fundamental constructs: the \emph{actor}, the \emph{technology}, the \emph{activity} and the \emph{software artefact}. These constructs are well--accepted in software engineering research~\cite{sjoberg2008building}. These constructs exist within a \emph{context} \cite{petersen2009context, dybaa2012works, clarke2012situational} which, by its nature, is difficult to define.

The scope of the theory, and therefore of the propositions, will align in some way with the empirical world being field-studied. For example, a researcher working with a theory and propositions about requirements engineering is unlikely to be empirically investigating code inspections.

Participants in a field study are expected to be drawn from the empirical world as that world aligns with the theory and propositions. Remaining with our example, the researcher would likely conduct field studies with requirements engineers as the participants, and ask those requirements engineers about requirements engineering. Also, the researcher will empirically study attributes of the four constructs as they relate to requirements engineering, e.g., requirements engineers (\emph{actor}), who use requirements gathering templates (\emph{technology}), to elicit (\emph{activity}) requirements (\emph{software artefact}), all within a \emph{context}.

\subsection{Participants as a sampling frame}

\begin{figure}
    \centering
    \includegraphics[scale=0.4]{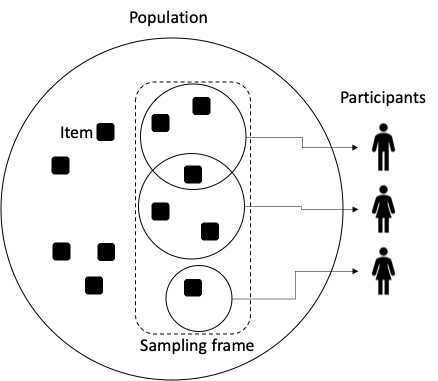}
    \caption{Participants as a sampling frame. Each participant has access to one or more items of interest from the population, and the aggregate of participants provides a sampling frame for sampling the items of interest.}
    \label{figure:participants-as-a-sampling-frame}
\end{figure}

In Section~\ref{subsection:sampling}, we discussed the sampling of items from a population of interest.
We suggest that, depending on the theory and propositions, it may be more effective to treat participants as a kind of sampling frame through which the items of interest from the empirical world are sampled, and therefore indirectly studied. Our suggestion is illustrated in Figure \ref{figure:participants-as-a-sampling-frame}.

Remaining with our earlier example, if the researcher intends to study the \emph{attitude} of requirements engineers then it makes sense to treat requirements engineers as the population and to sample from that population. This is because attitude is a property, or attribute, of the requirements engineers themselves. But if the researcher intends to study any one or more of the other three constructs of theories -- i.e., \emph{activity}, \emph{technology} or \emph{artefact} -- or \emph{actors} other than the participant, then the population of interest is not the requirements engineer but rather one or more of these other constructs. More strictly, the population of interest is likely to be a configuration of \emph{actor}, \emph{activity}, \emph{technology} and \emph{artefact}, all within one or more contexts. The researcher should ideally sample from across all of the appropriate constructs of interest. The implication is that participants should be recruited for the ``access'' they give to the population of \emph{actor}, \emph{activity}, \emph{technology} and \emph{artefact}, and not sampled for their representativeness as practitioners. In other words, a participant should be recruited for the contribution they can make to the formation of a sampling frame for sampling the items of actual interest. Recruiting participants in this way helps to ensure that the researcher collects information -- either facts about the world, or beliefs about the world -- that are drawn in relation to the items of interest.

\subsection{Participant roles}
\label{subsubsection:roles}

Participants will often be software practitioners who are located somewhere \emph{within} a software engineering situation and can therefore perceive other \emph{actors}, as well as the \emph{activities}, \emph{technologies} and \emph{artefacts} of the situation. Using Falessi et al.'s~\cite{falessi2018empirical} R\textsuperscript{3} model, discussed in Section~\ref{subsection:Rmodel}, practitioners would be expected to have some degree of Real, Relevant, and Recent experience. 

Sometimes a participant may be a software practitioner who is located outside of the situation but who can, for some reason, contribute to the formation of a sampling frame. Also, sometimes the practitioner has real and relevant experience, but the experience is not recent.

For our framework, we therefore define three roles for a participant in a field study:
\begin{itemize}
    \item The participant who is a \emph{Performer} within the situation, e.g., a programmer.
    \item The participant who is an \emph{Observer}, but not a \emph{Performer}, and is located elsewhere within the situation, e.g., a tester may observe aspects of the programmer's behaviour and performance.
    \item The participant who is an \emph{Advisor} with experience from a range of other, but related, situations. A common example here is a consultant who has not performed in the particular situation, or observed it, but draws on professional experience from elsewhere. The more experienced, and the more widely experienced, a software practitioner becomes the more likely they will have \emph{some} real, relevant experience, though that experience may not be recent.
\end{itemize}

Each of these roles provides information to the researcher, but this information is of different degrees of credibility, e.g., the information may be drawn from real experience but the experience, and therefore the information, may not actually be relevant. Similarly, the Performer will have experienced a situation contemporaneously (they did something at the time), but that experience may no longer be recent in relation to when the field study is being conducted. Later in this section, we map the three roles to the three elements of Falessi et al.'s~\cite{falessi2018empirical} R\textsuperscript{3} model.

A practitioner's experience may allow that practitioner to be classified as a \emph{Performer} for one study, whilst some other of the same practitioner's experience may allow that practitioner to be classified as an \emph{Observer} for another study, and some other of the same practitioner's experience may allow that practitioner to be classified as an \emph{Advisor} for yet another study; and indeed some of the practitioner's experience may not allow classification according to the three roles, e.g., because that experience is not relevant. For example, a software tester may be a Performer in a study of software testing, an Observer in study of programming, and an Advisor in a study of testing in another context. We acknowledge the complication of particular practitioners having multiple roles; it is a complication inherent with research into SE.

We also recognise the potential role of an \emph{expert} and intentionally do not use the word ``expert'' as a label for any of our roles. This is because a Performer or an Observer or an Advisor may be an expert, depending on their experience.

Our suggestion is that, when undertaking an empirical study, and when \emph{recruiting} participants, the researcher evaluates a practitioner against these three roles in the order that we have presented them, i.e., first determine whether the practitioner could be treated as a \emph{Performer}. According to our framework, a practitioner who could not satisfy any of the three roles would in principle be rejected as a participant in the study. In practice, the researcher may introduce an additional role or roles for practitioners, where it is appropriate to do so for their research. For example, researchers may introduce the role of \emph{Client} in a requirements engineering activity to recognise a participant who is able to make observations of the requirements engineer, but perhaps lacks relevant experience of requirements engineering to be an Observer.

\subsection{Practitioner characteristics affecting the information they provide}
\label{sec:characteristics}

In addition to the three participant roles, we suggest five characteristics of the practitioner that might affect the quality of information they can provide as a participant in a study. We identified these characteristics using the dialectic process briefly described in Section~\ref{subsection:framework-development}. The five characteristics are:


\begin{description}
    \item \textbf{Quantity of experience}: In general, the greater the quantity of situations experienced by the practitioner, the more experienced is the practitioner for the research. 
    \item \textbf{Perceptual sensitivity}: The more the practitioner is able to \emph{carefully} perceive events and, through that careful perception, to help prevent or reduce bias in their perception, the more valuable is the practitioner for the research. 
    \item \textbf{Situation selectivity}: The more the practitioner can distinguish between the different situations and so provide only the more real, relevant and recent information to the researcher, the more valuable is the practitioner for the research. For this characteristic, it may be more valuable to the researcher for the practitioner to \emph{restrict} the information they share to a specific situation, or situations, to ensure the information is more valid and more relevant, cf. the role of the sampling frame.
    \item \textbf{Reflexivity}: The more the practitioner is able to subsequently \emph{reflect} on those perceptions and, through that reflection, to help prevent or reduce biases in the information they share with the researchers, the more valuable is the practitioner for the research.
    \item \textbf{Willingness}: The practitioner must have the willingness or, alternatively phrased, the motivation to openly share information.
\end{description}

These five characteristics are all simplifications, and are all very challenging to measure. Again, the aim is to encourage researchers to think carefully about the recruitment of more credible participants, i.e., those practitioners who are able to provide better quality information to the research. 

\subsection{Demographics of participants}
\label{section:demographics}

For clarity, we distinguish between the demographics of participants and what might be called the demographics of the phenomenon of interest, e.g., the context of the actor-activity-technology-artefact configuration. Given our focus on recruiting credible participants, this subsection focuses on \textit{participant} demographics. In Section~\ref{section:reference-examples} we also recognise the demographics, or context, of the phenomenon of interest. 

Empirical studies of software engineering often report the demographics of the participants. For example, in one of their survey studies of professional software engineers at Microsoft, Begel and Zimmerman~\cite{begel2014analyze} found statistically significant differences in responses to 29 questions, based on the demographics of the respondents. And taking one pervasive demographic -- gender -- as another example, Carver and Serebrenik~\cite{carver2019} summarise several papers published in the 2019 edition of the International Conference on Software Engineering (ICSE) that discuss gender and software engineering.

These examples highlight the importance of considering demographics when collecting information from practitioners about their experience of software practice, i.e., practitioners with differing demographics may have differing experiences and may therefore provide different information. Also, it may be that the three roles we have identified have different demographic profiles, e.g., different proportions of gender or age for the different roles. Because of the inherent nature of demographics, and its relevance to software engineering research and practice, we explicitly recognise demographics in our framework, rather than only implying it in the existing characteristics.

Also, for clarification, we include the practitioner's \textit{functional role} (e.g., Systems Engineer, Programmer, or Tester) as part of the participant's demographics. We include functional role under demographics for a few reasons. We want to include functional role in the framework whilst also recognising that both functional roles and the \textit{titles} for functional roles can vary substantially from project to project, and from company to company etc. For these reasons, functional role is not necessarily a reliable indicator of the credible practitioner and is therefore not included in our five quality characteristics.

\subsection{A concise presentation of the framework}
\label{section:framework-summary}

A concise summary of the framework is presented in Table \ref{table:summary-of-framework}. As noted earlier, participants can provide two types of information to researchers, i.e., facts about the phenomenon of interest and beliefs about the phenomenon.  The table organises the participant roles, the five characteristics, the demographics, and the two types of information. Different roles can provide different types of information and, depending on the characteristics of the participant, different degrees of quality of information.  We intentionally do not try to populate the Quality criteria: these are complex criteria that are hard to measure and our primary interest in the current paper is to encourage researchers to consider these criteria when recruiting practitioners.

\begin{table*}[htp]
    \small
    \caption{A summary of the framework}
    \label{table:summary-of-framework}
    \centering
    \begin{tabular}{| l || c | c | c || c | c | c | c | c || c || c | c | }
        \toprule
        
        & \multicolumn{3}{ c ||}{\textbf{R\textsuperscript{3} model}} & \multicolumn{5}{ c ||}{\textbf{Quality}} & & \multicolumn{2}{ c |}{\textbf{Information}}\\
        \textbf{Role} & \textbf{Real} & \textbf{Rlvnt} & \textbf{Rcnt} & \textbf{E} & \textbf{P}  & \textbf{S} & \textbf{R} & \multicolumn{1}{ c ||}{\textbf{W}} & \textbf{D} & \textbf{Fact} & \textbf{Blfs} \\
        \midrule
        Performer & YES & YES & YES & \multicolumn{5}{c ||}{\dots} & \multicolumn{1}{c ||}{\dots} & YES & YES \\
        Observer & YES & yes & yes & \multicolumn{5}{c ||}{\dots} & \multicolumn{1}{c ||}{\dots}& yes & yes\\
        Advisor & yes & maybe & maybe & \multicolumn{5}{c ||}{\dots} & \multicolumn{1}{c ||}{\dots} & -- & yes\\
        \bottomrule
        \multicolumn{12}{l}{\textbf{Notes for R\textsuperscript{3} model:}}\\
        \multicolumn{12}{l}{\hspace{3mm} YES = strongly meets criterion; yes=moderately meets criterion}\\
        \multicolumn{12}{l}{\hspace{3mm} maybe = may meet criterion depending on context.}\\
        \multicolumn{12}{l}{\textbf{Notes for Quality criteria:}}\\
        \multicolumn{12}{l}{\hspace{3mm} We intentionally do not populate the Quality criteria.}\\
        \multicolumn{12}{l}{\hspace{3mm} E = Quantity of experience; P = Perceptual sensitivity; }\\
        \multicolumn{12}{l}{\hspace{3mm} S = Situation selectivity; R = Reflexivity; W = Willingness}\\
        \multicolumn{12}{l}{\textbf{Notes for Demographics (D):}}\\
        \multicolumn{12}{l}{\hspace{3mm} We explicitly recognise demographics because of its relevance to SE.}\\
        \multicolumn{12}{l}{\textbf{Notes for Information:}}\\
        \multicolumn{12}{l}{\hspace{3mm} YES = more likely to provide credible information}\\
        \multicolumn{12}{l}{\hspace{3mm} yes = may be able to provide credible information}\\
        \multicolumn{12}{l}{\hspace{3mm} -- = cannot provide facts about the item/s of interest}\\
    \end{tabular}
\end{table*}

\subsection{Additional considerations}
\label{section:additional-considerations}

When applying the framework some additional considerations will need to be addressed. As one example, focusing on the more credible participants might reduce the number of participants and therefore increase the likelihood that a given participant might be identifiable, e.g., when conducting an interview study in a company. This raises ethical issues. We do not consider these issues in this paper, however Strandberg~\cite{Strandberg2019}, for example, discusses ethical issues affecting interviews in software engineering. As a second example, \emph{finding} the more credible participants may be challenging, particularly as the researcher becomes more selective on the characteristics of the participant. But, as we noted from Bernard et al.'s~\cite{bernard1984problem} work in Section~\ref{section:accuracy}, collecting information from many inaccurate participants will not solve the accuracy problem, and will, as a consequence, also not produce relevant findings.

\subsection{Applying the framework: a brief example}

In Section \ref{section:reference-examples} we discuss three exemplars in much more detail. Here we present a simple example to illustrate aspects of the framework.

Returning to our earlier example, a researcher may invite requirements engineers to participate in a field study. The framework encourages the researcher to consider whether and how the invited practitioners provide access to the empirical world, as well as the quality of the information the participant can provide. Not all requirements engineers can provide the kinds of access the researcher may want, and not all requirements engineers can provide the quality of information the researcher may want. A requirements engineer (\emph{actor}) who is actively undertaking requirements engineering (\emph{activity}) using appropriate resources (\emph{technology}) and producing requirements specifications (\emph{artefacts}) that align with the theory and propositions of the research model is expected to have the most direct contact with the empirical world; in other words, may be understood as a \emph{Performer} who can, in principle, provide the most credible information to the researcher, e.g., facts about the world. Overall, the \emph{Performer's} information is expected to be more Real, Relevant and Recent (timely) compared to, for example, a requirements engineer who is an \emph{Observer}. The least credibility occurs with the requirements engineer who is an \emph{Advisor} as this participant has both the greatest variability in their experiences; but also, the researcher has the \textit{least certainty} in the Realness, Relevance and Recentness of the \emph{Advisor's} experience and information for the respective study. 

The three roles provide a convenient way of evaluating practitioners when recruiting participants for a study. These three roles support a coarser-grained evaluation, however; the five characteristics, discussed in sections \ref{sec:characteristics}, provide a finer-grained approach to evaluating practitioners for recruitment, but are much harder to implement.

By contrast, a common approach to recruiting participants for field studies in software engineering appears to based on the practitioner's functional role (e.g., project manager, software engineer, software tester) and their years of experience.

\subsection{Contribution of the framework}

One common approach to recruiting participants in field studies, such as interviews and surveys, is convenience sampling. With such an approach, practitioners are recruited based on some indicator or indicators of relevant experience, such as functional role and years of experience. Such an approach provides limited indication of the credibility of the respective practitioner and of the quality of information that the practitioner can provide.

In summary, the proposed framework encourages the researcher to think more explicitly in terms of the credibility of the practitioner and of the quality of information that the practitioner can provide. We propose three participant roles -- Performer, Observer, and Advisor -- and suggest a relative ranking of credibility to those roles, i.e., the Performer is the most credible practitioner, in relative terms. These roles provide a convenient way of evaluating practitioners when recruiting participants for a study. But the three roles only support a coarser-grained evaluation, albeit more finely grained than functional role and years of experience alone. We propose five other characteristics for ensuring the quality, i.e., quantity of experience, observational ability, situation selection, reflective ability, and willingness. We also recognise the importance of demographics. While these five quality characteristics are finer-grained, they are also much harder to evaluate. In Section~\ref{section:reference-examples} we show how elements of the framework have been used by others.

\section{Demonstrating the framework with exemplar studies}
\label{section:reference-examples}

\subsection{Overview}

In this section, we demonstrate how three exemplar articles~\cite{freimut05,singer1998practices,Vredenburg02} map to the elements of the framework. We take each exemplar in turn and then summarise all three exemplars, in the form of two tables, in the final subsection. In Section~\ref{section:method}, we explained how we selected these three exemplars.

Our mapping of the exemplars into the framework is clearly not exhaustive; it does not need to be. It is sufficient to show enough of a mapping to demonstrate that the articles, and the primary study reported by the respective article, contains enough elements of the framework. There are excerpts from each of the exemplar articles that we might report, and a given excerpt may be relevant to more than one element of the framework. Furthermore, a mapping of some part of an article into the framework is not an indicator of how substantially the article addressed the respective element, only an indicator that the respective researchers were aware of the element and took some action to address it. Similarly, the researchers of the respective study and article may not have considered an element to the same level of abstraction that we have in the framework.

As we show in the following subsections, all three exemplars provide support for the framework and its elements; furthermore, the three articles include insights on credible participants in field studies in SE that are not covered in the advice offered in the guidelines and other advisory sources. Thus, we argue that the three exemplars demonstrate the need for the framework, and also demonstrate that existing guidelines provide limited advice on the topic of recruiting credible practitioners for different forms of field study.

\subsection{Case study}
\label{sec:freimut}

The case study we consider was published by Freimut et al.~\cite{freimut05} as a journal article in 2005. The article proposes a model to measure cost-effectiveness of inspections, as well as a method to determine cost-effectiveness by combining project data and expert opinion. Expert opinion was gathered through interviewing 23 experts. The article is particularly relevant to our framework for two reasons: first, the article discusses the nature of expert opinion; second the article demonstrates many elements of the framework.

Before discussing the details of Freimut et al.'s article~\cite{freimut05} it is helpful to clarify two terms used in the article. First, Freimut et al.'s use of the term ``expert'' maps most closely to our use of the term ``Performer''. As briefly discussed in Section \ref{subsubsection:roles}, we consider that Performers, Observers and Advisors may each have a degree of expertise, e.g., that a practitioner might be a relatively novice Performer or, alternatively, an expert Performer. Second, Freimut et al.~\cite{freimut05} use the term ``role'' to refer to the \textit{functional} role of the practitioner, i.e., they distinguish between practitioners who are Developers, Analysts and Testers. In our framework, by contrast, the \textit{functional} role is an element of participant demographics; we reserve the term ``role'' in the framework to Performer, Observer or Advisor.

Turning to the details of the article, and first considering the nature of expert opinion, Freimut et al.~\cite{freimut05} devote a whole section of their article to discussing the nature of, and the need for, expert opinion. They recognise that one problem, for which expert opinion can help, is when information about a phenomenon cannot be collected by any other viable means, such as through measurement, observation or experimentation. Freimut et al.~\cite{freimut05} recognise that expert data is subject to bias, uncertainty and incompleteness, but also that these problems can be controlled through carefully performed elicitation of expert estimates. 

Furthermore, Freimut et al.~\cite{freimut05} discuss the selection of experts and present two criteria for selection:
\begin{enumerate}
    \item The \emph{role} of the expert in the development process. For Freimut et al., experts must have access to the information they are supposed to estimate. To do this, the experts must have participated in the respective process, e.g., in design inspections. For Freimut et al., \emph{performers} of a relevant task qualify as experts (according to their definition of expert) by virtue of being performers. They write, ``\dots for the purpose of effort estimation, \emph{people performing the corresponding tasks} qualify as experts.'' (\cite{freimut05}, p. 1081; emphasis added). 
   
    \item The level of expertise: Freimut et al. write, ``Such experience needs to be \textit{sufficiently varied and extensive with respect to the targeted tasks}.'' (\cite{freimut05}, p. 1081; emphasis added).

\end{enumerate}

In terms of demonstrating that the Freimut et al.~\cite{freimut05} study implicitly uses the elements of the framework, we organise the demonstration in terms of the main categories of the framework:

\begin{itemize}
    \item \textbf{Information:} In the case study, \emph{factual} information was collected as project data from the QA (Quality Assurance) management. The experts were asked to provide \emph{beliefs}, e.g., as probability distributions, about defects and costs.  Freimut et al. performed several analyses to assess the validity of those beliefs.
    
    \item \textbf{Quality}: As recognised in Section~\ref{section:framework-summary} of our article, the Quality criteria are hard to measure etc. There are some indications that Freimut et al. were at least implicitly thinking about some of these criteria. For example:
    
    \begin{itemize}
        \item \textbf{Quantity of experience:} this is indicated by the experts' years of experience.
        
        \item \textbf{Perceptual sensitivity:} for this criterion, there are no clear examples in the Freimut et al. article.
        
        \item \textbf{Situation selectivity:} as one example, Freimut et al. discuss how, in the pilot study, experts were accidentally lead to think of unusual instances rather than typical cases.
       
        \item \textbf{Reflexivity:} as one example, Freimut et al. discuss how an Analyst reflected on how they did not have recent experience. As another example, Freimut et al. discuss how testers in the pilot study unanimously agreed that it was possible for them to estimate the required parameters with their experience, though also recognised that some parameters would be more difficult to estimate.
       
        \item \textbf{Willingness:} Freimut et al. briefly discuss how experts were recruited on the basis of being proactive and motivated and how the recruited practitioners were ``highly motivated during the interviews'' and ``showed great interest in the study.''
    \end{itemize}
    
    \item \textbf{Demographics}: Freimut et al. do not explicitly discuss demographics; the one indication of demographics in their article is implicitly in terms of gender. When they refer to experts in their study they refer only to male experts, e.g., ``Similarly, an expert who is currently not performing the task about which \emph{he} is to provide an estimate may provide different estimates than an expert who is performing the task.'' (\cite{freimut05}, p. 1091; emphasis added).
    
    \item \textbf{The R\textsuperscript{3} model}: Freimut et al. provide a range of comments that clearly map to the R\textsuperscript{3} model. As examples:
    
    \begin{itemize}
        \item \textbf{Real}: The levels of expertise for the experts ranged from 3 to 18 years.
        
        \item \textbf{Relevant}: Freimut et al. discuss how a particular expert, who is responsible for a very complex part of the system, provides estimates that are significantly different from other experts who are responsible for less complex parts of the system. All of the experts have Real experience; here, the Relevance of the expert's opinion is based on the complexity of the part of the system on which the developer is working.
        
        \item \textbf{Recent}: Freimut et al. explain that one Analyst had not participated in analysis inspections for a long time and, because the Analyst's estimates differed from the other Analysts, Freimut et al. decided to exclude this estimate.
    \end{itemize}
   
    \item \textbf{Roles:} Freimut et al. identify three functional roles, i.e., Developer, Analyst and Tester. All three of these functional roles map to the role of Performer in the framework. Each functional role \emph{performs} a different activity in software development and is therefore appropriate for providing information on different aspects of the cost--effectiveness of inspections.
\end{itemize}

\subsection{Interview study}

The interview study we consider was published by Singer~\cite{singer1998practices} as a conference paper in 1998. Singer's article concerns the maintenance of large scale software systems. Information was collected from participants through paired interviews, i.e., two participants participated in each interview. The interview questionnaire comprised three parts: background information, task analysis, and a tools wish-list. Due to time constraints and other factors, the third part of the questionnaire was rarely asked during the interview. Participants were recruited by managers. Singer does not report how many participants were recruited but indicates that participants were drawn from ten industrial sites.

Singer's article is published very ``early'', relative to most guidelines in software engineering. Thus, the article does not use any (stated) guidelines when making research--design decisions. Considering exemplars published \textit{prior} to the publication of guidelines allows us to examine whether the elements of the framework were implicitly recognised by at least some researchers before guidelines were established, and therefore whether the framework concerns recurring, more fundamental issues. 

As with the Freimut et al.~\cite{freimut05} article, Singer~\cite{singer1998practices} also considers the nature of ``experts'', although not in the systematic way undertaken by Freimut et al. Singer writes, on the basis of the background information she collected, that, ``These data paint a picture of software maintenance engineers as being both \emph{expert} programmers and \emph{experts} in the project in which they are working.'' (\cite{singer1998practices}). The article later further clarifies that the experts were both expert programmers and expert maintenance programmers. In terms of the R\textsuperscript{3} model, this suggests participants whose expertise is Relevant to programming, to maintenance programming, and to the application domain/project.

Once again, we organise our demonstration of the mapping of the Singer~\cite{singer1998practices} article around the main categories of the framework:

\begin{itemize}
    \item \textbf{Information:} The first part of the interview questionnaire collected  contextual information about software maintenance (i.e., applications, languages, platforms and projects). This information may broadly be understood as collecting factual information. By contrast, the four, qualitative ``truths'' proposed by Singer were based on \textit{beliefs} gathered through the second part of the questionnaire that concerned task analysis.
   
    \item \textbf{Quality}: There are some indications that Singer was at least implicitly thinking about some of these criteria. For example:
    
    \begin{itemize}
        \item \textbf{Quantity of experience:} The Singer article reports information (min., max. and mean) on years experience, time on programming language, number of languages and time on project.
        
        \item \textbf{Perceptual sensitivity:} for this criterion there are no clear examples in the Singer article.
        
        \item \textbf{Situation selectivity:} for this criterion there are no clear examples in the Singer article, however Singer notes that, ``The [ten corporate] environments themselves were diverse with respect to practically all defining variables.''
        
        \item \textbf{Reflexivity:} One reason that Singer designed the study to interview participants in pairs was to encourage ``\dots them to verbalize their thoughts because they could talk to each other about aspects of the project/product.'' Whilst Singer has not explicitly referred to reflection, the exchange of information between the two participants would, at least in principle, encourage reflection.
        
        \item \textbf{Willingness:} Singer writes, ``\dots it is possible that the managers chose their more stable employees to participate in the interviews.'' and that she used the paired--interview design to make the situation more comfortable for the participants. Whilst these two examples do not explicitly relate to willingness to share information, they would, at least in principle, encourage participants to share information.
    \end{itemize}
    
    \item \textbf{Demographics}: The article refers to collecting ``background information''. This information relates to the projects and not to the participants. It appears therefore that no participant--related demographic information was collected.
    
    \item \textbf{The R\textsuperscript{3} model}: Singer provides a range of comments that clearly map to the R\textsuperscript{3} model. As examples:
    
    \begin{itemize}
        \item \textbf{Real}: All participants had a minimum of three years experience, with a minimum of one year on the project.
        
        \item \textbf{Relevant}: Singer observed that approx. 60\% of the software maintainers' professional life was spent on maintenance projects and approx. 40\% on new development. From this she speculates, ``It is not clear if different skills are needed for these two endeavors [maintenance vs new development], but if so, then, on average, the interviewees were more familiar with the job of maintaining software programs than developing new ones.'' In terms of our framework, the participants' experience would therefore be more Relevant to maintenance than to new development.
        
        \item \textbf{Recent}: Singer writes that all participants had to be working on a product that was at least 1.5 years old and ``currently in a maintenance phase.''
    \end{itemize}
    
    \item \textbf{Roles:} Singer writes, in relation to the managers selecting the participants, ``It was stressed to the managers that all participants should be involved in the actual maintenance of software (as opposed to leading a team, other administrative posts, etc.).'' Singer was therefore selecting Performers.
\end{itemize}

\subsection{Survey study}

The survey study we consider was published by Vredenburg et al.~\cite{Vredenburg02} as a conference paper in 2002. Vredenburg et al.'s article concerns the use of methods, practices, key factors and trade-offs for user-centred design (UCD). The survey questionnaire was distributed to attendees at the CHI'2000 conference and then via email to members of the Usability Professional Association (UPA). 103 participants completed the survey questionnaire.

Like the Friemut et al.~\cite{freimut05} article and the Singer~\cite{singer1998practices} article, Vredenburg et al.~\cite{Vredenburg02} also consider experts, however Vredenburg et al.~\cite{Vredenburg02} refer to the participants as ``opinion leaders''. They write, ``They [the participants] were likely opinion leaders in the UCD community, playing a leading role in their own organization’s UCD practice.''

Two particularly interesting aspects of the survey are, first, the way that the study recruited participants, and second, the results of the survey. For the first aspect, Vredenburg et al.~\cite{Vredenburg02} defined a target participant (``at least three years of experience with UCD, and considered UCD as their primary job.'') and highlighted in the invitation--to--participants that only those who met the target profile should participate. Vredenburg et al. therefore encouraged prospective participants to self--select, or self--reject, themselves. Also, Vredenburg et al.~\cite{Vredenburg02} asked participants to consider a representative project. Vredenburg et al.~\cite{Vredenburg02} were therefore looking to recruit participants with at least Real and Relevant experience, as well as participants who were \emph{Performers} in the situation of interest. Also, Vredenburg et al.~\cite{Vredenburg02} are distinguishing between the participant and the item of interest, e.g., the representative project.

For the second aspect, concerning the results, we consider two examples here. First, Vredenburg et al. observe a \emph{lack of consensus} in the responses: the 103 participants identified a total of 191 indicators of UCD effectiveness. The lack of consensus, and therefore the amount of ``disagreement'', suggests that cultural norms were not influencing the responses. Second, whilst participants identified UCD practices that were considered useful, they were rarely used: ``Only three of the [top 10] measures [for UCD success] were reported by more than 10\% of the respondents and none of them was higher than 20\%.'' This is curious for it suggests that \textit{Performers} are not performing the practice.

As with our two other exemplars, we organise our demonstration of the mapping of the Vredenburg et al.~\cite{Vredenburg02} article around the main categories of the framework:

\begin{itemize}

    \item \textbf{Information:} The article provides information about project profiles (e.g., number of people on the team). Such information may broadly be understood as \emph{factual}. By contrast, the information on, for example, measures of UCD effectiveness and applied measures, were based on \emph{beliefs} gathered from participants who took part in the (representative) project/s.
    
    \item \textbf{Quality}: As recognised in Section \ref{section:framework-summary} of our article, the Quality criteria are hard to measure etc. There are some indications that Vredenburg et al. were at least implicitly thinking about some of these criteria. For example:
    
    \begin{itemize}
        \item \textbf{Quantity of experience:} The Vredenburg et al. article collected information on years of experience with UCD, percentage of work time on UCD-related activities over the past 12-months, number of projects involving UCD over the past 12 months, and level of familiarity with UCD practices. 
        
        \item \textbf{Perceptual sensitivity:} For this criterion there are no clear examples in the Vredenburg et al. article.
        
        \item \textbf{Situation selectivity:} Participants were asked to select a representative project that used UCD, and in which they had participated, over the past 12 months.
        
        \item \textbf{Reflexivity:} For this criterion there are no clear examples in the Vredenburg et al. article.
        
        \item \textbf{Willingness:} For this criterion there are no clear examples in the Vredenburg et al. article although, presumably, completion of the questionnaire survey is an indicator of at least some willingness to share information.
    \end{itemize}
   
    \item \textbf{Demographics}: The study collected limited information: country in which the participant worked, and on highest qualification (e.g., PhDs or Masters).
    
    \item \textbf{The R\textsuperscript{3} model}: Vredenburg et al. provide a range of comments that clearly map to the R\textsuperscript{3} model. As examples:
    
    \begin{itemize}
        \item \textbf{Real}: Vredenburg et al. write, ``\dots respondents appeared to be truly experienced practitioners because of their multiple years of experience and familiarity with UCD, and the fact that they attended the CHI conference or were members of the UPA.''
        
        \item \textbf{Relevant}: Vredenburg et al. asked the participants to choose a representative project. They observed that nearly 63\% of the respondents chose an Internet/Intranet project.
        
        \item \textbf{Recent}: As noted above, the article collected information on number of projects, and percentage of work time, involving UCD over the most recent 12 months. Vredenburg et al. found that on average (mean and mode) participants participated in five projects involving UCD. 
    \end{itemize}
    
    \item \textbf{Roles:} Vredenburg et al. write that participants were asked to select a representative project that used UCD, \emph{and in which they had participated}, suggesting that participants were either Performers or Observers.
\end{itemize}

\subsection{Summary of exemplars}

Table \ref{table:mapping-exemplars-to-framework} and Table \ref{table:compare-exemplars-with-framework} summarise the mappings from the three exemplar articles to the framework. Table \ref{table:mapping-exemplars-to-framework} provides more detail on the mapping, at least for some of the elements of the framework, whilst Table \ref{table:compare-exemplars-with-framework} provides a concise mapping that directly aligns with Table~\ref{table:summary-of-framework} in Section~\ref{section:framework-summary}.

As noted earlier, all exemplars demonstrate support for the framework; furthermore, the exemplars include insights on credible participants in field studies in SE that are not covered in the advice offered in the guidelines and other advisory sources. Thus, we argue that the three exemplars demonstrate the need for the framework, and also demonstrate that existing guidelines provide limited advice on the topic of recruiting credible practitioners for different forms of field study.

\begin{table*}[htp]
    \small
    \caption{Summary of exemplar articles mapping to the framework}
    \label{table:mapping-exemplars-to-framework}
    \centering
    \begin{tabular}{| l |  c | c | c |}
        \toprule
        & \multicolumn{3}{c |}{\textbf{Exemplars}}\\
        \textbf{Criterion} & \textbf{Freimut~\cite{freimut05}} & \textbf{Singer~\cite{singer1998practices}} & \textbf{Vredenburg~\cite{Vredenburg02}}\\
        \midrule
        \multicolumn{4}{| l |}{\textbf{Meta-information about the study}}\\
        Year & 2005 & 1998 & 2002 \\
        Topic & Inspections & Maintenance & User centred design\\
        Study type & Case study & Interview & Survey \\
        Sample & 23 people & 10 sites & 103 people\\
        Medium & Journal & Conference & Conference \\
        \midrule
        \multicolumn{4}{| l |}{\textbf{Information}}\\
        Beliefs & Estimates & ``Truths'' & Methods etc\\
        Facts & Costs & Background & Project profiles\\
        & & information &\\
        \midrule
        \multicolumn{4}{| l |}{\textbf{Quality criteria}}\\
        Experience & Yes & Yes & Yes \\
        Perception & None & None & None\\
        Selection & Some & None & Yes \\
        Reflection & Some & Some  & None \\
        Willingness & Yes & Some & Some\\
        \midrule
        \multicolumn{4}{| l |}{\textbf{Demographics}}\\
        Demographics & No & No & Some \\
        \midrule
        \multicolumn{4}{| l |}{\textbf{Falessi et al.'s~\cite{falessi2018empirical} R\textsuperscript{3} model}}\\
        Real & Yes & Yes & Yes \\
        Relevant & Yes & Yes & Yes \\
        Recent & Yes & Yes & Yes\\
        \midrule
        \multicolumn{4}{| l |}{\textbf{Role}}\\
        Performer & Developers (7) & Maintainers & Probably\\
         & Analysts (n/16) & & \\
         & Testers (m/16) & & \\
        Observer & None & None & Possibly\\
        Advisor & None & None & None\\
        \bottomrule
    \end{tabular}
\end{table*}

\begin{table*}[htp]
    \small
    \caption{Comparison of exemplar articles with the framework}
    \label{table:compare-exemplars-with-framework}
    \centering
    \begin{tabular}{| l || l || c | c | c || c | c | c | c | c || c || c | c | }
        \toprule
        
        \textbf{Art-}& \textbf{Role} & \multicolumn{3}{ c ||}{\textbf{R\textsuperscript{3} model}} & \multicolumn{5}{ c ||}{\textbf{Quality}} & &  \multicolumn{2}{ c |}{\textbf{Info.}}\\
        
        \textbf{icle} & \textbf{POA} & \textbf{Real} & \textbf{Rlvnt} & \textbf{Rcnt} & \textbf{E} & \textbf{P}  & \textbf{S} & \textbf{R} & \textbf{W} & \multicolumn{1}{| c ||}{\textbf{D}} & \textbf{Fact} & \textbf{Blfs} \\
        \midrule
        1 \cite{freimut05} & P - - & Y & Y & Y & Y & N & y & y & Y & N & Y & Y\\
        2 \cite{singer1998practices} & P - - & Y & Y & Y & Y & N & N & y & y & N & y & Y\\
        3 \cite{Vredenburg02} & p o - & Y & Y & Y & Y & N & Y & N & y & y & y & Y\\
        \bottomrule
        \multicolumn{13}{l}{\textbf{Notes:}}\\
        \multicolumn{13}{l}{Y = There is clear evidence that the article maps, for this element.}\\
        \multicolumn{13}{l}{N = There is no evidence that the article maps, for this element.}\\
        \multicolumn{13}{l}{y = There is some evidence that the article maps, for this element. }\\
        \multicolumn{13}{l}{P = Performer; O = Observer; A = Advisor ; - = None}\\
        \multicolumn{13}{l}{p = probably Performer; o = possibly Observer}\\
        \multicolumn{13}{l}{E = Quantity of experience; P = Perceptual sensitivity; S = Situation selectivity}\\
        \multicolumn{13}{l}{R = Reflexivity; W = Willingness; D = Demographics}\\
        \multicolumn{13}{l}{}\\
    \end{tabular}
\end{table*}
\section{Conclusion}
\label{section:conclusion}

Based on our authorship of books on case study research~\cite{runeson2012book} and controlled experiments~\cite{wohlin2012experimentation}, we hypothesised that advice on \emph{recruiting credible participants} for field studies is limited. Our perception was, for example, that existing guidelines and checklists did not adequately advise on the recruitment of credible participants, in contrast to advising on the importance of sampling from a population. 

To investigate our hypothesis, we reviewed existing guidelines, checklists and other advice on a range methods of field study, e.g., specific guidelines on case studies, interviews, surveys, focus groups, and ethnography, and more general guidelines on recruiting participants, on empirical studies and empirical standards. Our review corroborated our hypothesis, i.e., while guidelines recommend the recruitment of appropriate participants, the guidelines provide little detail on how to identify credible participants for the respective study.

We then developed a framework to help researchers determine credible practitioners. We demonstrated the framework using three exemplar primary studies, i.e., a case study, an interview study and a survey study. Each of the exemplar studies implicitly or explicitly recognised some, but not all, of the characteristics identified in the framework.

In terms of our two research questions: for RQ1 (i.e., ``What guidance is currently available on the recruitment of credible participants for field studies of software engineering?''), we reviewed published guidelines, checklists and advisory information in software engineering research; and for RQ2 (i.e., ``How might we determine a credible participant?''), we proposed and demonstrated a framework.

The exemplar studies (i.e.,~\cite{freimut05,singer1998practices,Vredenburg02}) show that at least some prior research has at least implicitly recognised some of the issues we include in our framework, and sought to address these issues in their respective studies; and the guidelines explicitly recognise the importance of sampling participants. In contrast to sampling participants from a population, the proposed framework explicitly identifies a set of characteristics for researchers to consider when recruiting practitioners as credible participants for their field studies. More fundamentally, however, the proposed framework is intended to encourage researchers to think strategically about how they recruit participants for their empirical studies.

We recognised briefly, in Section~\ref{section:additional-considerations} that there are limitations to the framework and opportunities for further research, e.g., whilst the framework identifies characteristics, these characteristics may not be easy to apply.

In summary, the framework is intended to support researchers when conducting field studies in software engineering with human participants, particularly in research studies such as case studies, interviews and surveys.















\bibliographystyle{elsarticle-num}
\bibliography{references.bib}







\end{document}